\documentclass[lettersize,journal,nosections]{IEEEtran}
\IEEEoverridecommandlockouts

\usepackage[switch]{lineno}  
\usepackage{xcolor}

\usepackage{cite}
\usepackage{amsmath,amssymb,amsfonts}
\usepackage{threeparttable}
\usepackage{multirow}
\usepackage{booktabs} 
\usepackage{algorithmic}
\usepackage{graphicx}
\usepackage{textcomp}
\usepackage{svg}
\usepackage{tcolorbox}
\usepackage{colortbl}   
\usepackage{booktabs}  
\usepackage{listings}       
\usepackage{xcolor}         
\usepackage[utf8]{inputenc} 
\usepackage{textcomp}       
\usepackage{beramono}       
\usepackage{caption} 
\definecolor{keywordpurple}{RGB}{170,0,120} 
\definecolor{codegreen}{rgb}{0.8588, 0.9922, 0.9255}   
\definecolor{codestring}{rgb}{0.5,0,0.5}    
\definecolor{codegray}{rgb}{0.5,0.5,0.5}    
\newcommand{\lstbg}[3][0pt]{{\fboxsep#1\colorbox{#2}{\strut #3}}}

\lstdefinestyle{VerilogNormalStyle}{
    language=Verilog,                   
    basicstyle=\ttfamily\scriptsize,  
    linewidth=\columnwidth,
    xleftmargin=1em,
    framexleftmargin=1em,
    xrightmargin=0pt,     
    keywordstyle=\color{keywordpurple}\bfseries, 
    commentstyle=\color{codegreen},     
    stringstyle=\color{codestring},     
    identifierstyle=\color{black},      
    numbers=left,                       
    numberstyle=\tiny\ttfamily\color{codegray}, 
    numbersep=5pt,                      
    captionpos=b, 
    showspaces=false,                   
    showstringspaces=false,             
    showtabs=false,                     
    frame=tb,                           
    framesep=3pt,                       
    rulesepcolor=\color{black},         
    tabsize=4,                          
    breaklines=true,                    
    postbreak=\mbox{\quad},
    breakatwhitespace=true,             
    inputencoding=utf8,                 
    literate=
    {_}{\_}{1}
    {~}{{\texttildelow}}1
    {<=}{{$<=$}}1
    {>=}{{$\ge$}}1
    {==}{{$=$}}1
    {!=}{{$\neq$}}1
    {&&}{{$\land$}}1
    {||}{{$\lor$}}1
    {|}{{\textbar}}1
    {&&&}{{$\land\land\land$}}1
    {|||}{{$\lor\lor\lor$}}1
    {<<}{{$\ll$}}1
    {>>}{{$\gg$}}1
    {->}{{$\rightarrow$}}1
    {|}{{\textbar}}1,
    morecomment=[f][\color{red}]{---}, 
    morecomment=[f][\color{codegreen}]{+++},
    morecomment=[f][\lstbg{red!20}]{-\ },
    morecomment=[f][\lstbg{codegreen}]{+\ },
    morecomment=[f][\color{blue}]{@@},
}

\definecolor{mygreen}{RGB}{192,255,192} 
\definecolor{myred}{RGB}{255,204,204}   
\definecolor{lightgray}{RGB}{230,230,230} 

\def\BibTeX{{\rm B\kern-.05em{\sc i\kern-.025em b}\kern-.08em
    T\kern-.1667em\lower.7ex\hbox{E}\kern-.125emX}}
\begin{document}

\title{Rethinking LLM-Based  RTL Code Optimization Via Timing Logic Metamorphosis\\


}

\author{Zhihao Xu, Bixin Li$^{\ast}$~\IEEEmembership{Member,~IEEE,}, Lulu Wang,

\thanks{Z.Xu, B.Li, L.wang, are with the school of Computer Science and Engineering, Southeast University, Nanjing, China. E-mail: Hosea.xu@seu.edu.cn, bx.li@se.edu.cn, wanglulu@seu.edu.cn}

\thanks{*Corresponding author}
}

\maketitle

\begin{abstract}
Register Transfer Level(RTL) code optimization is crucial for achieving high performance and low power consumption in digital circuit design. However, traditional optimization methods often rely on manual tuning and heuristics, which can be time-consuming and error-prone. Recent studies proposed to leverage Large Language Models(LLMs) to assist in RTL code optimization. LLMs can generate optimized code snippets based on natural language descriptions, potentially speeding up the optimization process. 
However, existing approaches have not thoroughly evaluated the effectiveness of LLM-Based code optimization methods for RTL code with complex timing logic. To address this gap, we conducted a comprehensive empirical investigation to assess the capability of LLM-Based RTL code optimization methods in handling RTL code with complex timing logic.
In this study, we first propose a new benchmark for RTL optimization evaluation. It comprises four subsets, each corresponding to a specific area of RTL code optimization, \textit{logic operation optimization}, \textit{data path optimization}, \textit{timing control flow optimization}, and \textit{clock domain optimization}. Then we introduce a method based on metamorphosis to systematically evaluate the effectiveness of LLM-Based RTL code optimization methods.
Our key insight is that the optimization effectiveness should remain consistent for semantically equivalent but more complex code. After intensive experiments, we revealed several key findings.
(1) LLM-Based RTL optimization methods can effectively optimize logic operations and outperform existing compiler-based methods. 
(2) LLM-Based RTL optimization methods do not perform better than existing compiler-based methods on RTL code with complex timing logic, particularly in timing control flow optimization and clock domain optimization. This is primarily attributed to the challenges LLMs face in understanding timing logic in RTL code. 
Based on these findings, we provide insights for further research in leveraging LLMs for RTL code optimization.
\end{abstract}


\section{Introduction}

Optimizing Register Transfer Level (RTL) code is a critical step in the early stages of circuit design~\cite{R37,R38,R39,R41}. This process involves iteratively rewriting original RTL code snippets into optimized versions based on optimization patterns or synthesis feedback. Traditionally, this process heavily relies on the expertise of seasoned engineers~\cite{RTLRewriter,R37}. However, the increasing complexity of design patterns significantly hampers the efficiency of manual optimization. In contrast, existing compiler-based methods have limited scope and suboptimal performance in optimizing complex designs, and they also fall short in leveraging synthesis feedback to refine the code~\cite{symrtlo}. 

\begin{figure}[!t]
    \centering
    \includegraphics[width=0.99\linewidth]{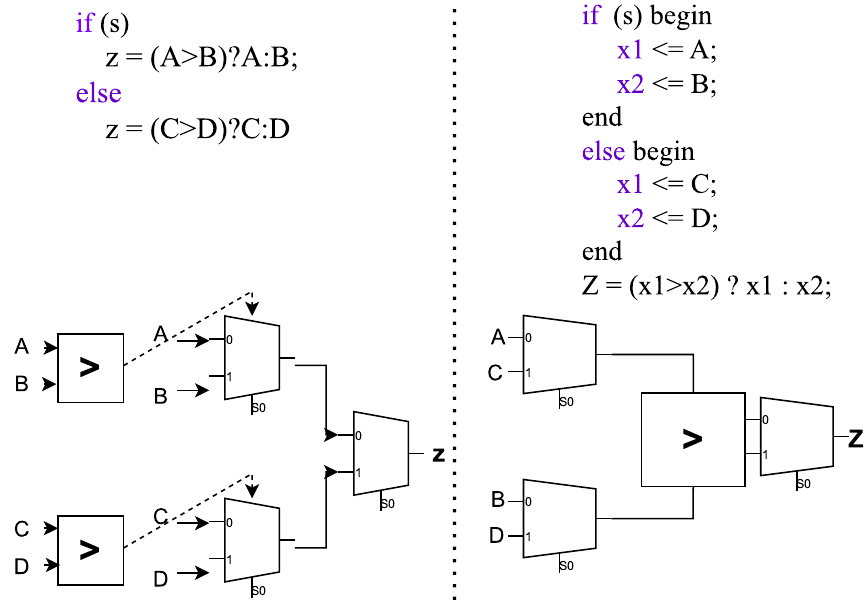}
    \caption {Example of LLM-Based Mux optimization of RTL code}\label{fig:1}
\end{figure}

Thus recent studies have proposed the use of Large Language Models (LLMs) to assist in RTL code optimization~\cite{RTLRewriter}. LLMs can generate optimized code snippets based on natural language descriptions, potentially speeding up the optimization process. For example, \figurename~\ref{fig:1} shows a classic example of LLM-Based Mux optimization of RTL code.
In this example, the LLM-Based code optimization method moves conditional comparison operations as far forward as possible to occur before the MUX selection. This reduces redundant computations by operational units (e.g., comparators, adders),  lowering the resource utilization and power consumption of the actual circuit.
Existing open-source compilers, such as yosys~\cite{yosys}, struggle to effectively handle such scenarios~\cite{RTLRewriter,symrtlo}.

However, existing LLM-Based RTL optimization methods only focus on optimizing the logic operations and data flow of RTL code like data paths, Muxes, and control signals~\cite{RTLRewriter}. They do not consider the timing characteristic of RTL code, even though it can directly impact the performance, power consumption, and reliability of the digital circuits~\cite{R36,R37,R38}. For instance, poorly optimized timing logic can lead to series issues such as setup and hold violations, increased latency, and functional failures in multi-clock domain RTL code. 
Therefore, it is essential to conduct a thorough investigation to identify the advantages and limitations of LLM-Based RTL code optimization methods when dealing with circuits that exhibit complex timing logic. The findings would illuminate domain-specific challenges and inform the design of more efficient LLM-based optimization methodologies.


To address this gap, we introduced a metamorphosis method to evaluate the effectiveness of LLM-Based RTL code optimization methods in handling complex RTL code, particularly those with intricate timing logic. And we also publish a new benchmark for RTL optimization. Specifically, we first collected RTL code samples from several existing benchmarks~\cite{Veval,rtllm} and open-source projects on GitHub, and organized them into a new benchmark by experienced RTL code developers. 
Then for systematically evaluate the capability of LLM-based RTL code optimizers across diverse RTL design scenarios, we categorize RTL optimization into four key dimensions:\textit{logic operation optimization}, \textit{data path optimization}, \textit{timing control flow optimization}, and \textit{clock domain optimization}. .
These four categories are reflected in FPGA vendor guidelines, EDA tool documentation, and industry best practices~\cite{chen2009design,sasao1993logic,zimmermann2005rtl,synopsys2011dc,doulos2020fsm,amd2023ug949,cadence2014cdc,vemeko2025fpga,unnikrishnan2024rtl} provide a structured and comprehensive basis for evaluating semantic robustness of optimization under varying RTL complexity.
For each area, our method designs a metamorphosis strategy to evaluate the effectiveness of LLM-Based RTL code optimization methods in this area. 
These metamorphosis strategies aim to generate new RTL code which is semantically equivalent to the original code but with increased complexity, particularly in timing logic.
We refer to these cases as mutant RTL code.
Our key insight is that the optimization effectiveness should remain consistent for semantically equivalent RTL code. 
Therefore, for evaluation we used different RTL code optimization methods include LLM-Based RTL code optimization methods to the optimize original RTL code and its mutant RTL code. 
Finally, we compared the optimization effectiveness of these methods on the original RTL code and the mutant RTL code.

Through intensive experiments we found that although LLM-Based RTL code optimization methods have shown great potential in optimizing logic operations and data paths, they do not perform better than existing compiler-based methods on RTL code with complex timing, particularly in timing control flow optimization and clock domain optimization. Then we conducted a comprehensive analysis to identify the reasons behind this limitation. Our contributions are as follows:
\begin{itemize}
    \item We introduced a metamorphosis method to evaluate the effectiveness of LLM-Based RTL code optimization methods in handling complex RTL code, particularly those with intricate timing logic.
    \item To our best knowledge, we are the first to conduct a comprehensive empirical investigation to assess the capability of LLM-Based RTL code optimization methods in handling complex timing logic. 
    \item We provided a new benchmark to evaluate the effectiveness of RTL Code optimization method to optimize RTL code with intricate timing logic.

\end{itemize}

\section{Background and Motivation} 
In this section, we first introduce the background of RTL code and its special characteristics of timing logic. Then we present the motivation for our work with a simple example.

\subsection{RTL Code}

Descriptions at the RTL level typically employ hardware description languages (HDLs) such as Verilog or VHDL. The core distinction of RTL code, compared to conventional software code, lies in its inherently sequential and synchronous temporal semantics~\cite{SIMTAM,LegoHDL}.In software programming languages (such as C, Java, or Python), instructions typically execute sequentially and variables are updated instantaneously within the thread of execution. In contrast, RTL code expresses system behavior as a set of concurrent processes that evolve synchronously in response to a global timing reference, typically described as clock cycles. For example, as shown in Listing~\ref{lst:2.1} the RTL code implement a simple counter.
\begin{lstlisting}[style=VerilogNormalStyle, caption= Example of Metamorphosis strategies in Logic Operation Optimization , label={lst:2.1}]
module counter(
    input wire clk,
    input wire reset,
    output reg [3:0] count);
    always @(posedge clk or posedge reset) begin
        if (reset)
            count <= 4'b0000;
        else if (count == 4'b1111)
            count <= 4'b0000;
        else
            count <= count + 1'b1;
end
endmodule
\end{lstlisting}
Unlike in traditional programming languages, the counter value only changes on the rising edge of the clock (posedge clk). Specifically, when the reset signal is active, the counter is synchronously reset to 0. When the count reaches 15 (\texttt{4'b1111}), it resets to zero in the next clock cycle; otherwise, it increments by 1. Therefore, unlike traditional oftware programming languages optimizing, RTL code optimization requires careful consideration of timing logic. Not only is what happens important, but when it happens is crucial to RTL code.



\subsection{Illustrative Examples}

Existing LLM-Based RTL code optimization methods rely on the understanding capabilities of LLMs for RTL code and additional search mechanisms~\cite{RTLRewriter}. However, the ability of LLMs to comprehend RTL code, particularly its timing logic, has not been thoroughly validated~\cite{verilogeval,verigen,rtlfixer}.
To illustrate this, we present an example to demonstrate the challenges LLMs face in understanding timing logic in RTL code and explain how this motivates our work.
Existing studies often compare the embedding vectors of models to quantify the semantic understanding capabilities of LLMs~\cite{Pei,R46}. 
Given a program snippet \( P \), a language model encodes it into a continuous vector representation \( \mathbf{E}(P) \in \mathbb{R}^d \), where \( d \) is the dimension of the model's latent embedding space.
A common assumption in embedding-based program analysis is that the semantic similarity between programs correlates with the distance between their embeddings, the closer the embeddings, the more similar the semantics~\cite{Pei}.





To evaluate the capability and sensitivity of large language models (LLMs) in understanding RTL code, particularly its temporal logic, we conduct a structured study inspired by the principles of mutation testing. Specifically, we first generated numerous initial test cases using the latest RTL code generators~\cite{LegoHDL}. Subsequently, we designed four types of semantically equivalent mutations and four types of semantic-breaking mutations. By comparing the embedding vectors distance between the original and mutated RTL snippets, we assess how well LLMs capture the semantic structure of RTL, especially with respect to timing behavior.

\begin{figure}[!t]
    \centering
    \includegraphics[width=0.85\linewidth]{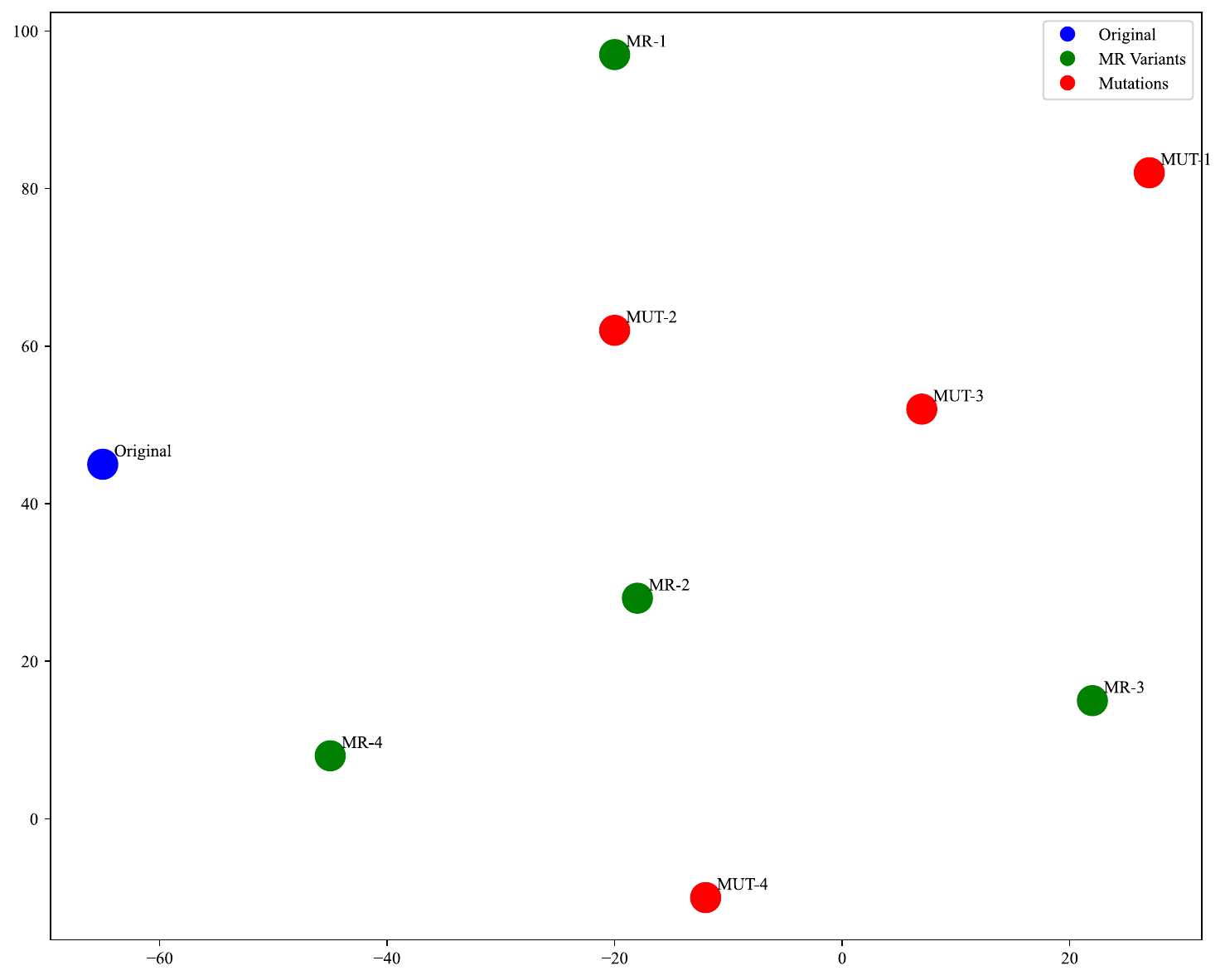}
    \caption {Embedding results of original and mutated code on CodeLlama}\label{fig:4}
\end{figure}

The semantically preserving mutations include: restructuring nested conditional logic into equivalent flattened forms, injecting redundant but behaviorally neutral logic, modifying the structure of time-sensitive logic by rewriting \texttt{always} blocks with equivalent control flow, and refactoring between syntactically different but semantically identical assignment styles. The semantic-breaking mutations include: inverting logical conditions to alter control flow, introducing minor temporal inconsistencies in sequential logic, modifying numeric thresholds or constants that control execution behavior, and introducing incorrect variable updates.

\begin{figure*}[!t]
\centering
 \includegraphics[width=0.80\textwidth]{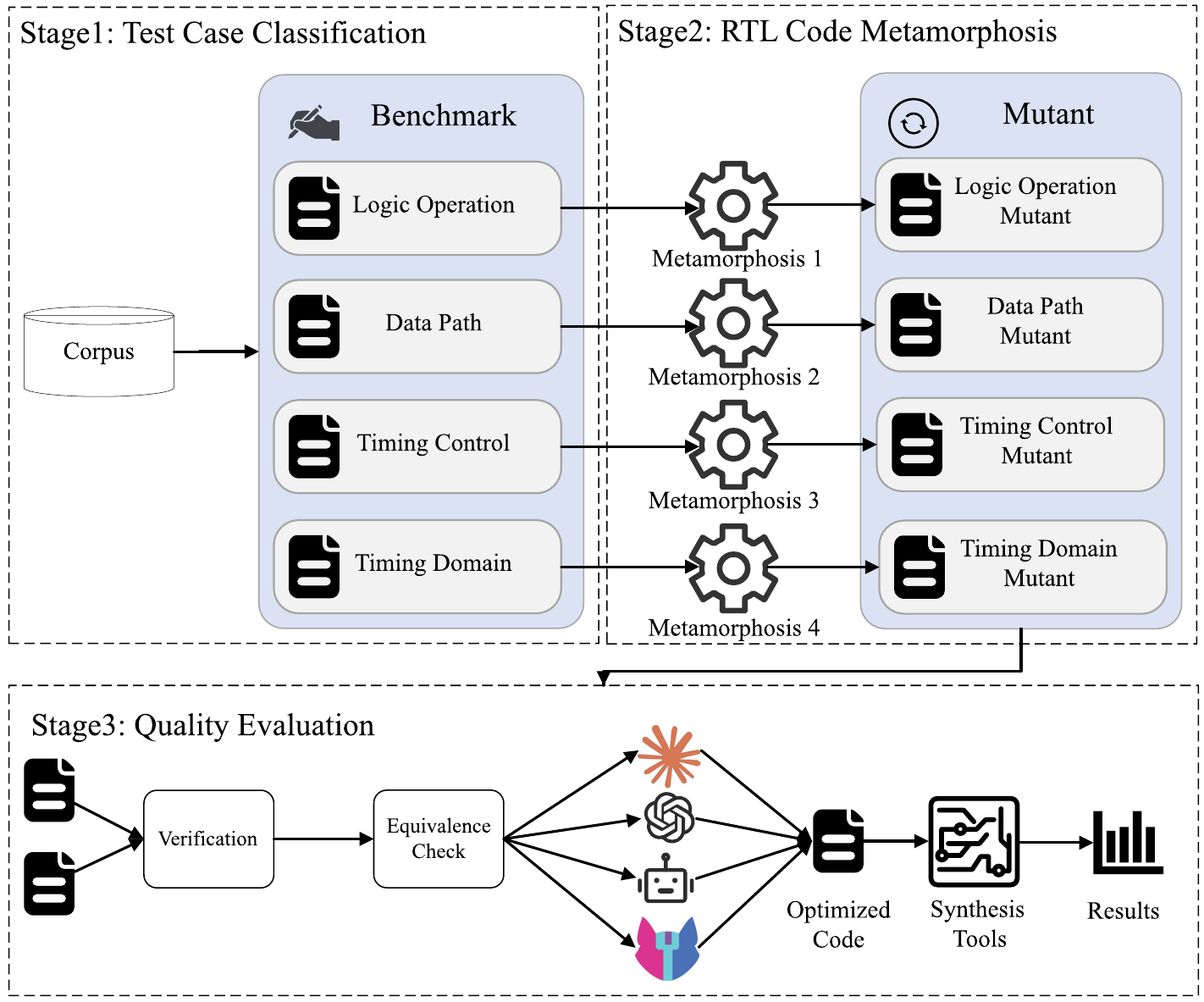}
\caption{Workflow of our Method to evaluate LLM-Based RTL Optimization Method}
\label{fig:workflow}
\end{figure*}

\figurename~\ref{fig:4} shows the embedding results of the original and mutated code on \textit{CodeLlama-7B}~\cite{codellama}. We use t-SNE to project high-dimensional embedding vectors into a two-dimensional space, allowing a more intuitive visualization of how different program variants are represented by the language model. In this space, the distance between points reflects the Euclidean distance between their embeddings—shorter distances indicate higher semantic similarity, while longer distances suggest greater semantic divergence. The blue point represents the original program (Original), green points denote our semantically preserving mutations (MR RTL code), and red points indicate mutated versions with altered semantics.

Notably, the distance between the MR-3 group and the original program is greater than that of other semantically preserving variants, even exceeding some semantic-breaking variants. Although MR-3 remains semantically consistent with the original program, its transformations of timing logic may have introduced deviations that the model failed to comprehend. Therefore, this inspired us to conduct a comprehensive investigation to determine whether LLMs can assist engineers in optimizing RTL code with complex timing logic.

\section{Methodology}

In this section, we introduce our methodology to evaluate the effectiveness of LLM-Based RTL code optimization methods in handling complex RTL code, particularly those with intricate timing logic. As shown as \figurename~\ref{fig:workflow}, our method includes three stages; Test Case Classification, RTL Code Metamorphosis, and Quality Evaluation. Specifically, we first collect existing benchmark RTL code datasets and classify them into four key areas of RTL code optimization in the stage of Test Case Classification. Then, we apply metamorphosis strategies to the original benchmark test cases to generate new functional equivalence test cases(Which called Mutants) in stage of RTL Code Metamorphosis. 
Finally, we evaluate the effectiveness of LLM-Based RTL code optimization methods by comparing the effectiveness of original RTL code and mutant RTL code in stage of Quality Evaluation.
Our method is inspired by existing code symmetry evaluation methods~\cite{R44,R45}. We adopted the same idea of metamorphosis to systematically evaluate the effectiveness of LLM-Based RTL code optimization methods in handling complex RTL code, particularly those with intricate timing logic.

\subsection{Test Case Classification}

To construct a benchmark of RTL Code optimization, we first collected RTL code in existing benchmarks of RTL code~\cite{Veval,rtllm} and Github. Then we invited a number of domain experts to classify the collected RTL code into four key areas of RTL code optimization, \textit{logic operation optimization}, \textit{data path optimization}, \textit{timing control flow optimization}, and \textit{clock domain optimization}. 
Through multiple rounds of screening and evaluation, we ensure that the RTL code included in each area is functionally non-overlapping and representative.
Thus, our benchmark is composed by four different parts, each part contains a set of RTL code snippets that are representative of the specific area of optimization.

\subsection{RTL Code Metamorphosis}

In this stage, we apply specific metamorphosis strategies to the four key areas of RTL code. These metamorphosis strategies are introduced as follows.


\subsubsection{Metamorphosis strategies of Logic Operation Optimization}

For the logic operation optimization, we employ a metamorphosis strategy based on Boolean algebra laws. The strategy applies a sequence of algebraic transformations to change the original structure of RTL code logic operation expressions while keep the semantic.

For example, let \( E = (a \land b) \lor (a \land c) \) represent the original expression. We first decompose \( E \) into its conjunction sub-expressions \( (a \land b) \) and \( (a \land c) \), then individually transform these sub-expressions using De Morgan's laws into their NOR equivalents \( \neg(\neg a \lor \neg b) \) and \( \neg(\neg a \lor \neg c) \), respectively. Subsequently, we recombine these transformed sub-expressions into a single nested expression, converting the outer disjunction into a conjunction structure. Thus the original expression \( E \) is transformed into a more complex form as shown as equation~\ref{eq:1}.
\begin{equation}\label{eq:1}
E \Rightarrow \neg\left[\neg\left(\neg(\neg a \lor \neg b)\right) \land \neg\left(\neg(\neg a \lor \neg c)\right)\right]
\end{equation}

\begin{lstlisting}[style=VerilogNormalStyle, caption= Example of Metamorphosis strategies in Logic Operation Optimization , label={lst:strategy1}]
    module logic_transformed(
        input a, b, c,
        output y
        );
-        assign y = (a & b) | (a & c);
+        wire term1 = ~(~a | ~b);
+        wire term2 = ~(~a | ~c);
+        wire term3 = ~(~term1 & ~term2);
+        wire term4 = ~(~(a & b & c));
+        assign y = ~(~term3 & ~term4);
    endmodule
  \end{lstlisting}

Listing \ref{lst:strategy1} illustrates the transformation process. Starting from a simple Boolean expression, we iteratively introduce redundant terms to systematically increase the syntactic complexity while preserving the original logical functionality. Each step replaces basic logical operations with more intricate forms and adds logically redundant sub-expressions, resulting in a functionally equivalent but complex Boolean expression.

\subsubsection{Data Path Optimization}

For data path Optimization, we propose a cascade data path metamorphosis strategy. This strategy increases data-flow complexity and diversifies multiplexer structures by decomposing direct assignments and selection logic into multi-stage, cross-referenced configurations.

Specifically, we first identify the main signals in the RTL code and construct the condition expressions that are always true based on these signals. Then, we randomly select insertion points to add conditional branches, using the always true condition expressions as the branch predicate. The original functional code is no longer placed directly within sequential or combinational logic blocks, but is instead migrated to the true branch of the conditional statement. Meanwhile, the else branch contains randomly generated logic expressions. These additions are never executed and do not affect the original RTL code semantics. By doing so, the originally clear execution path becomes encapsulated within unnecessary conditional branches to increase the static complexity of RTL code.

\begin{figure}[!t]
    \centering
    \includegraphics[width=0.90\linewidth]{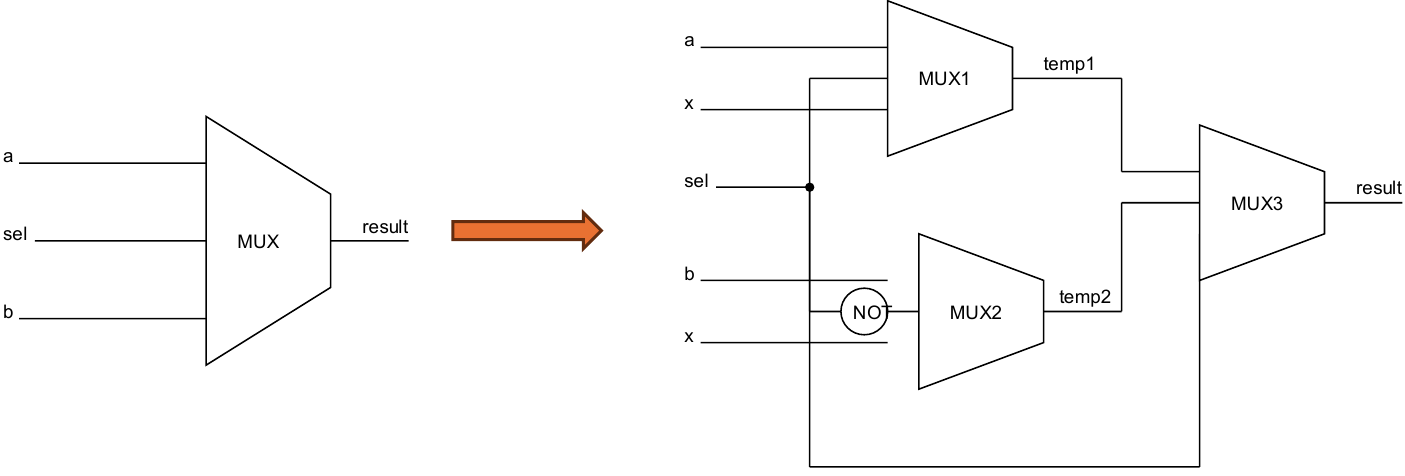}
    \caption{Examples of Metamorphosis strategies in Data Path Optimization}
    \label{fig:mux_refactor}
\end{figure}

Second, for multiplexer structures we convert standard multiplexers into nested or cascaded forms. For example, as shown in the left part of \figurename~\ref{fig:mux_refactor}, the RTL code implements a basic two-to-one multiplexer. We employ the data path metamorphosis strategy to change the original RTL code as shown in the right part of \figurename~\ref{fig:mux_refactor}. To increase the number of selection steps, we introduce an intermediate variable \(X\). The circuit now first selects between the two inputs and \(X\), and then produces the final output. This transformation does not change the original semantics of the RTL code and introduces redundant data path. By doing so, we introduce additional layers of selection logic, increasing the complexity of the multiplexer structure while preserving its original functionality.


\subsubsection{Timing Control Flow Optimization}

For timing control flow optimization, we propose a state machine metamorphosis strategy to evaluate the ability of LLM-Based RTL optimization method to recognize and simplify complex timing logic. 
The state machine metamorphosis strategy increases the complexity of finite state machines by transforming direct state transitions and monolithic functional states into more elaborate, multi-stage forms. Instead of allowing transitions to occur directly between two states, or executing multiple operations within a single state, the strategy distributes the transition or operation process across several intermediate or subdivided states. Specifically, the strategy firstly extends a simple state change, such as a transition from \( S_1 \) to \( S_2 \), into a chained sequence of pass-through states. For example, the direct transition is replaced by a longer path \( S_1 \to S_{1A} \to S_{1B} \to S_{1C} \to S_2 \), where each intermediate state does not introduce additional logic but merely propagates the system forward. 
Secondly, a single state that originally performs multiple operations will be subdivided into a series of sub-states, with each sub-state responsible for a part of the original logic. For instance, an operation performed in state \( S_1 \), such as updating \( A \) and \( B \) and then transitioning to \( S_2 \), can be split into two sub-states. The first sub-state updates \( A \) and transitions to the next sub-state. The second updates \( B \) and proceeds to \( S_2 \). This transformation retains functional equivalence while increasing the complexity of the control flow.

\begin{figure}[!t]
    \centering
    \includegraphics[width=0.70\linewidth]{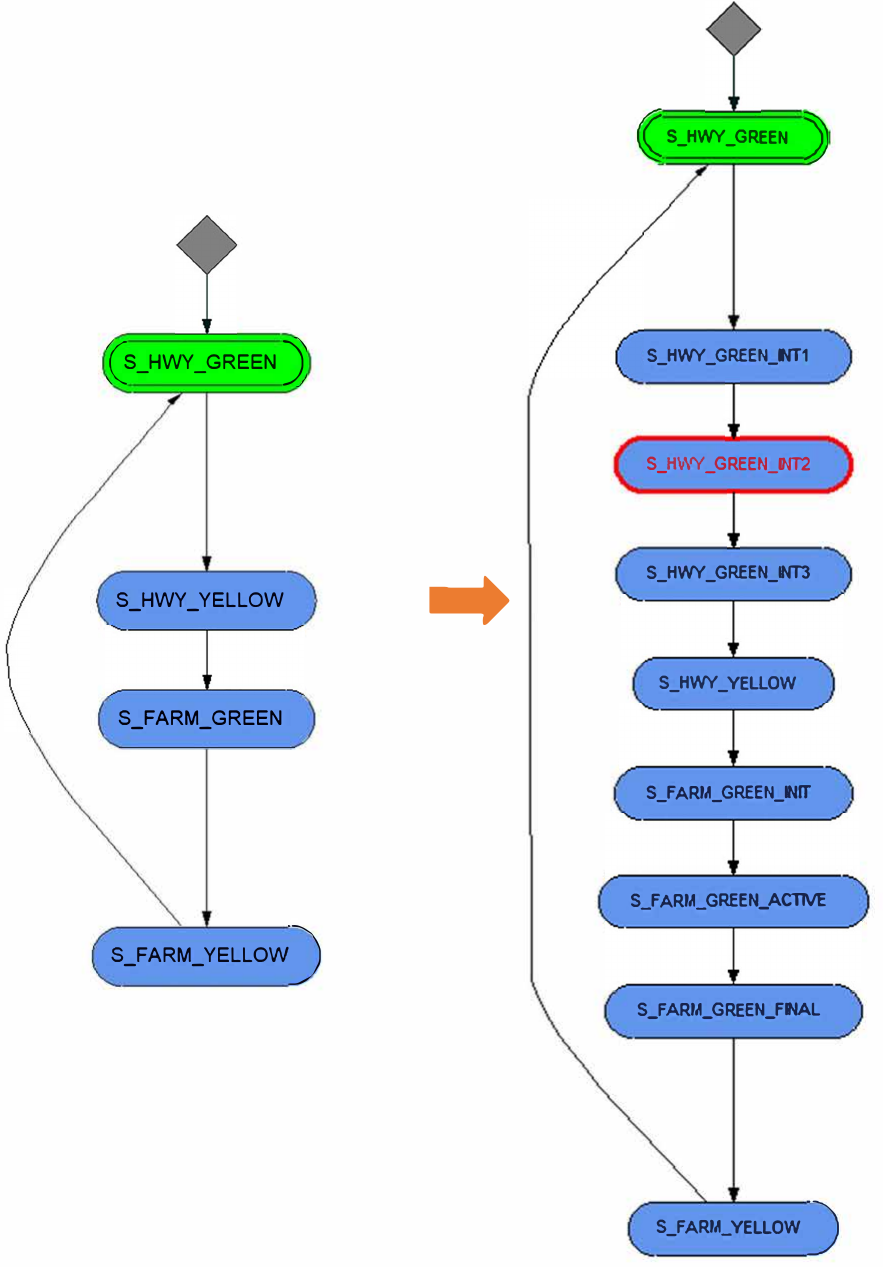}
    \caption {Example of Timing Control Flow Optimization Metamorphosis strategy}\label{fig:7}
\end{figure}

As shown as \figurename~\ref{fig:7}, we demonstrate how the state machine metamorphosis strategy is applied to the RTL code implementation of a traffic signal controller. Specifically, the left part of the figure illustrates the state transition diagram of the original RTL code, while the right part shows the state transition diagram of the transformed code.
In the original code, the highway green light state (\texttt{HWY\_GREEN}) transitions directly to the highway yellow light state (\texttt{HWY\_YELLOW}) after its timer expires. In the mutant version, we inserted three intermediate states to form a chained path. These intermediate states have no functional logic and simply transition to the next state, ultimately reaching \texttt{HWY\_YELLOW}. While the number of states has increased, the functionality remains equivalent.
Furthermore, the green light state is a single state lasting for 6 clock cycles. In the mutant RTL code we subdivided it into three functionally equivalent substates, initialization (\texttt{FARM\_GREENINIT}), active (\texttt{FARM\_GREEN\_ACTIVE}), and finalization (\texttt{FARM\_GREEN\_FINAL}). Together, these substates achieve the same functionality as the original state, with a total duration still spanning 6 clock cycles.

By doing so, we increase the complexity of the state machine (expanding from 4 states to 8 states) and introduce intricate transition paths. However, the overall functionality and behavior are fully consistent with the original code.

\subsubsection{Clock Domain Optimization}

For clock domain optimization, we propose a clock domain cross-structure metamorphosis strategy to evaluate the ability of LLM-Based RTL optimization method. 
The clock domain metamorphosis strategy first partitions and reconstructs existing clock domains. For example, a continuous logic chain typically operate under a single clock will deliberately split across two or more clock domains. The split clocks are chosen to have identical frequencies or a fixed multiple relationship to the original clock domains. Although the logic is functionally unchanged, new clock domain crossings and synchronizers are inserted into the original RTL code.

Then we change the standard CDC synchronizers into more intricate forms.  For example, we increase complexity by changing simple, standard clock domain crossing synchronization mechanisms into more convoluted asynchronous protocols. These changes do not alter the  original semantics of the RTL code, as the functional behavior and data integrity across clock domains are preserved. By doing so, we increase the clock domain heterogeneity of the original RTL code.

\begin{lstlisting}[style=VerilogNormalStyle, caption= Example of Metamorphosis strategies in Logic Operation Optimization , label={lst:strategy4}]
-        always @(posedge clk) begin
-           regA <= ...;          
-           regB <= comb(regA);  
-        end
+        always @(posedge clk1) begin
+           regA <= ...;                     
+           sync_reg1 <= comb1(regA);       
+        end
+        always @(posedge clk2) begin
+           sync_reg2 <= sync_reg1;   
+           regB <= comb2(sync_reg2);  
+        end
\end{lstlisting}

As shown as Listing \ref{lst:strategy4}, we apply the clock domain cross-structure metamorphosis strategy to a simple RTL code with time clock domain. In original RTL code, data is propagated from register \texttt{regA} to register \texttt{regB} through combinational logic, all operating under a single clock domain (\texttt{clk}).
Then, we split the logic such that \texttt{regA} and \texttt{regB} reside in different but synchronous clock domains (\texttt{clk1} and \texttt{clk2}), and we insert additional synchronization and combinational stages between them. The clocks \texttt{clk1} and \texttt{clk2} are set to have identical frequencies or a fixed multiple relationship, so there is no actual clock drift between domains.
Although additional synchronizer registers and combinational logic are introduced, the data is still deterministically and reliably transferred from \texttt{regA} to \texttt{regB}, preserving the original functional semantics of the circuit. This change does not affect the original functionality but increases the complexity of the clock domain.

\subsection{Quality Evaluation}

After applying metamorphosis strategies to the original benchmark RTL code, we obtain a set of new benchmark RTL code, which we call Mutants. Then we use different RTL code optimization methods to optimize the original benchmark RTL code and mutant RTL code. To ensure the semantic equivalence of the original benchmark RTL code and mutant RTL code, we perform formal verification and simulation to ensure that the original benchmark RTL code and mutant RTL code are semantically equivalent. Specifically, we use ABC~\cite{yosys} as our solver, which is highly efficient in circuit verification, and Icarus Verilog~\cite{Iverilog} as our simulator.
Then we synthesize the optimized RTL code of original benchmark RTL code and mutant RTL code using synthesis compiler to get gate-level circuit descriptions. Theoretically, 
the original RTL code and mutant RTL code should exhibit the same optimization results, as the metamorphosis does not alter the program semantics of the RTL code. Therefore, by comparing the gate-level circuit descriptions of original benchmark RTL code and mutant RTL code, we can evaluate the effectiveness of RTL code optimization methods in RTL code optimization.

\section{Evaluation}
In this section,  four experiments are conducted to evaluate the effectiveness of LLM-Based RTL code optimization methods in handling complex timing logic. Specifically, our evaluation aims at answering the following Research Questions(RQs).

\begin{itemize}
    \item RQ1: How effective are LLM-Based RTL code optimization methods in optimizing logic operations?
    \item RQ2: How effective are LLM-Based RTL code optimization methods in optimizing data paths?
    \item RQ3: How effective are LLM-Based RTL code optimization methods in optimizing timing control flow?
    \item RQ4: How effective are LLM-Based RTL code optimization methods in optimizing clock domains?
\end{itemize}

\subsection{Baseline}

We selected RTLRewriter~\cite{RTLRewriter} as one of our baselines.
RTLRewriter optimized RTL code using LLMs to rewrite the RTL code by partitioning the RTL code into smaller segments and using Retrieval Augmented Generation to strengthen the effectiveness of LLMs in optimizing RTL code.
It is the latest and most effective LLM-based method for RTL code optimization. 
And we selected GPT-4 and Claude-3.7-sonnet as the baseline models for comparison in our experiments, since they are state-of-the-art large models for code-related tasks. Our baseline method~\cite{RTLRewriter} also uses these two LLMs as its primary comparison baselines. Furthermore, we included Yosys, the most widely used open-source compiler, the same as our baseline method~\cite{RTLRewriter}.

\subsection{Benchmark}

Our benchmark suite was built upon two sources, the latest RTL code in Github and the existing benchmark~\cite{Veval,rtllm}, which includes basic patterns, data-path, memory, MUX, FSM, and control logic. This suite includes 54 RTL code cases related to logic operations, 27 cases related to data paths, 40 cases related to timing control flow, and 32 cases related to clock domains. 

\subsection{Implementation}

In this research, we implemented our metamorphosis strategies using Python and Verilog. Our evaluation ran on a computer with an Intel Core i9 CPU@2.10GHZ, 128GB RAM and an NVIDIA A6000 GPU using CUDA 12.0. We evaluated the performance of LLM-Based RTL code optimization methods by five metrics, wires, cells, area, delay, and power. These metrics are commonly used in RTL code optimization and synthesis, and they are defined as follows. 
\begin{itemize}
    \item Wires: represent the interconnections in a RTL code. A lower number of wires indicates a more efficient RTL code.
    \item Cells: represent the logical elements utilized in the RTL code, where a higher count signifies greater logical complexity. 
    \item Area: represents the physical space occupied by the RTL code. It includes  the number of logic gates, flip-flops, and
    interconnections required to implement the RTL code. A smaller area indicates a more compact and efficient RTL code.
    \item Delay: represents the time it takes for a signal to propagate through the RTL code. A lower delay indicates a more efficient RTL code.
    \item Power: represents the power consumption of the RTL code. A lower power consumption indicates a more energy-efficient RTL code.
    
\end{itemize}

\subsection{Prompt Design}

Our prompt design is informed by the baseline approach~\cite{RTLRewriter}, which also employs a large language model to perform RTL code optimization under natural language instructions. In their framework, the LLM is prompted with a general instruction to optimize RTL code while preserving functionality, without extensive prompt engineering or task-specific demonstrations. Following this paradigm, we adopt a fixed zero-shot prompt across all models and experiments to eliminate prompt variability as a confounding factor. 

Specifically, our prompt takes the form as follows. 

\begin{center}
    \begin{tcolorbox}[colback=gray!10,
                      colframe=black,
                      width=\linewidth,
                      arc=1mm, auto outer arc,
                      boxrule=0.5pt,
                     ]
    [Prompt Template]
    You are an expert hardware engineer optimizing RTL code for performance and area. 
    Below is a RTL module. Please optimize it while keeping its functionality unchanged.

    [RTL Snippet]

    module foo (...);\\
    ...\\
    endmodule

    \end{tcolorbox}
\end{center}

This instruction is immediately followed by the RTL code snippet. The output is expected to be a valid, semantically equivalent optimized RTL code. Our prompting strategy align with that of the baseline and applying it consistently across all mutation scenarios. By standardizing the prompt across all experiments, we ensure fair comparison while isolating the LLM’s intrinsic capability in RTL code optimization from the influence of prompt variability.
 More advanced prompt tuning and adaptation are left as future work, and we discuss their potential impact in Section~\ref{sec:threats}.

\subsection{Metric Normalization}

To the effectiveness of LLM-based RTL optimization, we require a reference implementation that serves as the functional and structural baseline. In our study, we use the output of the industry-standard toolchain Yosys as the de facto ground truth. Specifically, for each input RTL design, we synthesize it using Yosys with standard optimization passes enabled (e.g., opt, techmap, abc) to obtain a canonicalized, optimized version. This synthesized result is treated as the reference against which we compare the output of LLM-based optimization methods. Since Yosys is a mature, widely adopted open-source synthesis tool that applies deterministic and semantics-preserving optimizations, it provides a stable and reproducible target for evaluation.
Thus, we report all quantitative metrics (e.g., wire count, cell count, area, delay, power) as \textbf{normalized ratios} relative to the \textbf{Yosys}.
For a given metric $M$ and RTL code instance $d$, the normalized ratio under a method $\mathcal{A}$ is computed as follows.
\[
\text{Ratio}_{\mathcal{A}}^{(d)} = \frac{M_{\mathcal{A}}^{(d)}}{M_{\text{Yosys}}^{(d)}}
\]
We then compare the \textbf{mean normalized ratio} of each method across RTL code for each key area.
The normalization allows fair comparisons across different RTL code by evaluating the relative performance change from baseline.

\subsection{Answer to RQ1}

\begin{figure}[!t]
    \centering
     \includegraphics[width=\linewidth]{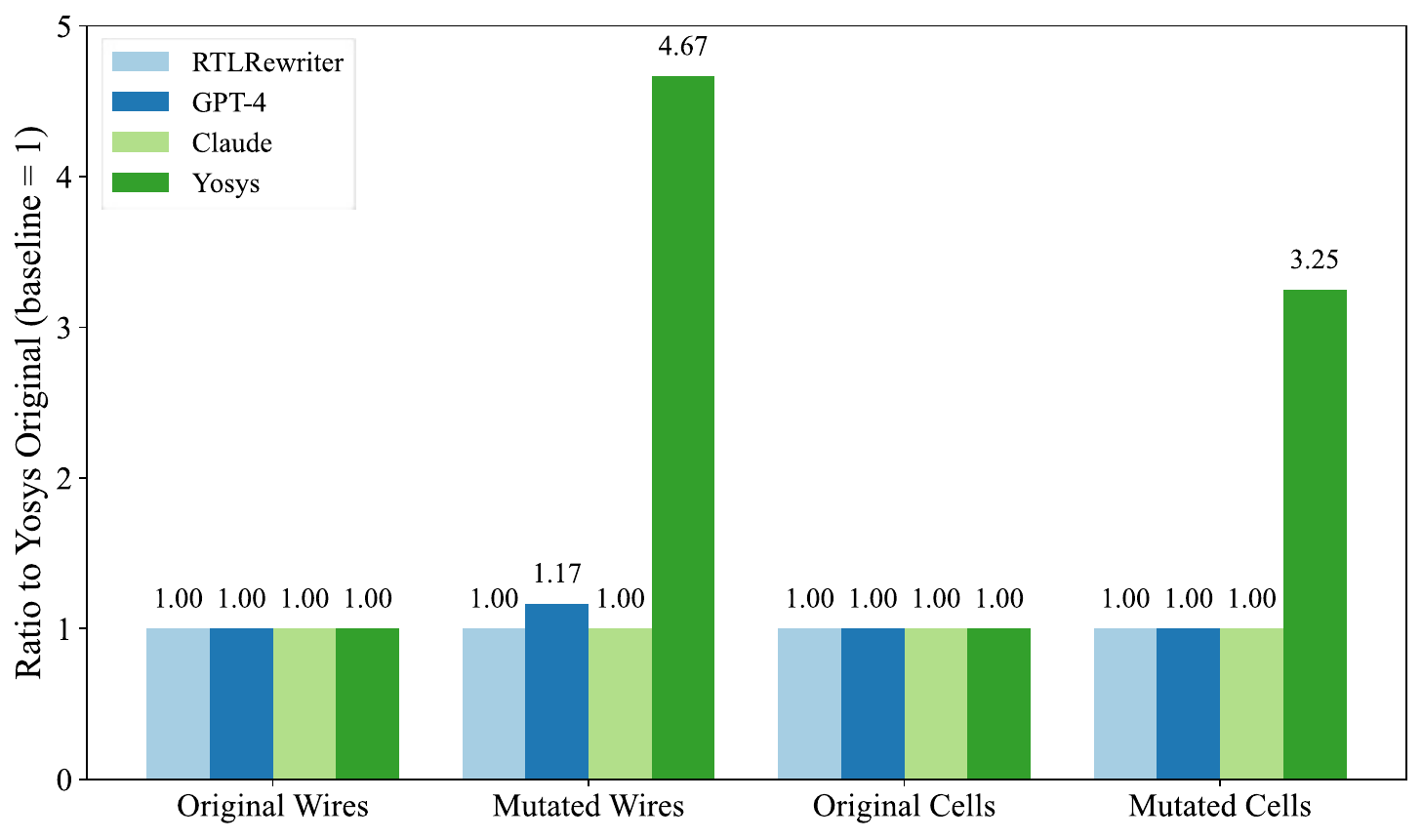}
    \caption{Logic Operation optimization Results with Wires and Cells}
    \label{fig:exp1}
\end{figure}

As shown in \figurename~\ref{fig:exp1}, we present the optimization results on wires and cells for various RTL code optimization methods targeting logic operation circuits. On the mutant RTL code, both \textbf{RTLRewriter} and \textbf{Claude-3.7-sonnet} achieved identical results to those on the original RTL code (wire and cell ratios remaining at 1.00× baseline), demonstrating strong robustness in logic operation optimization. In contrast, \textbf{GPT-4} showed a slight performance degradation, with the number of wires on the mutant RTL increasing to 1.17× baseline, while cells remained unchanged. This degradation primarily arises from the tendency of GPT-4 to employ suboptimal wire assignment strategies within the \texttt{always} block during code optimization. 
As shown as Listing \ref{lst:verilog_vA}, \textbf{GPT-4} employed continuous assignment for computations, which results in the creation of separate hardware structures.
Consequently, additional intermediate signals must be synthesized by the compiler during the hardware translation phase, resulting in increased circuit complexity.
\begin{lstlisting}[style=VerilogNormalStyle, caption=RTL Code Optimization Example of GPT, label={lst:verilog_vA}]
    wire [7:0] sum_result = a + b;
    wire [7:0] diff_result = a - b;
    wire [7:0] and_result = a & b;
    wire [7:0] or_result  = a | b;
  
    always @(*) begin
        case (op_code)
+            2'b00: result = sum_result; 
+            2'b01: result = diff_result; 
+            2'b10: result = and_result;  
+            2'b11: result = or_result;    
            default: result = 8'b0;        
        endcase
    end
\end{lstlisting}
\textbf{Yosys} exhibited the worst optimization performance on the mutant RTL code, with the number of wires and cells rising sharply to 4.67× and 3.25× baseline, respectively. It did not achieve reductions comparable to those observed on the original RTL code. This is primarily because Yosys struggles to handle multi-level selection logic and highly nested conditional expressions effectively.

\begin{center}
    \begin{tcolorbox}[colback=gray!10,
                      colframe=black,
                      width=\linewidth,
                      arc=1mm, auto outer arc,
                      boxrule=0.5pt,
                     ]
    \textbf{Overall.} The results indicate that LLM-Based RTL optimization methods are effective in optimizing logic operations, even in the presence of complex transformations.
    \end{tcolorbox}
\end{center}

\subsection{Answer to RQ2}

\begin{figure}[!t]
    \centering
     \includegraphics[width=\linewidth]{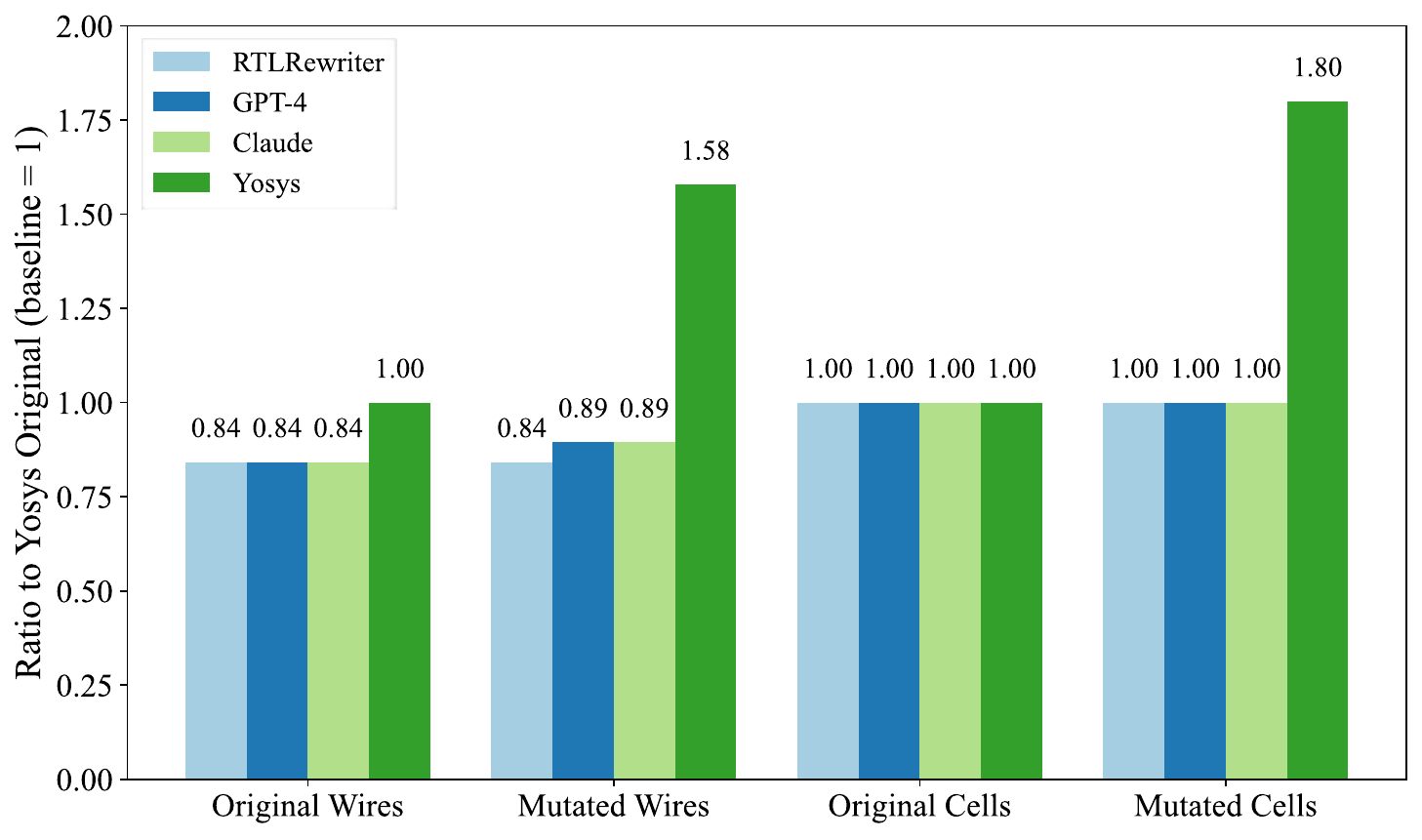}
    \caption{Data Path Optimization Results with Wires and Cells}
    \label{fig:exp2}
\end{figure}


The results, as shown in \figurename~\ref{fig:exp2}, indicate that all LLM-based methods and \textbf{RTLRewriter} outperform \textbf{Yosys} in optimizing both original and mutant RTL code with complex data path and multiplexer structures. For example, on the mutant RTL code, all LLM-based methods and RTLRewriter reduced the number of wires to 0.89–0.94× the baseline, while Yosys only achieved a ratio of 1.59. Similarly, the number of mutant cells for LLM-based methods and RTLRewriter remained at 1.00× baseline, whereas Yosys reached 1.80×.

Due to its reliance on abstract syntax trees (ASTs) of data path for circuit partitioning and optimization, RTLRewriter is especially effective at optimizing RTL code with intricate data path and MUX structures. \textbf{Claude-3.7-sonnet} and \textbf{GPT-4} also demonstrated superior performance compared to Yosys in data path optimization, but their effectiveness on mutant RTL code did not fully match their performance on the original RTL code. Upon further analysis, we found that in the absence of explicit instructions, LLMs may retain the non-executing or redundant code segments in the mutant RTL code. However, if we prompted with explicit directives to eliminate non-executing code, both Claude and GPT-4 would achieve optimization results comparable to those of RTLRewriter.

In contrast, Yosys showed limitations when dealing with complex data path and MUX structures. Its compiler-based optimizations lack the semantic reasoning capacity of LLMs and thus struggle to disentangle deeply nested or cross-referencing signals, resulting in a higher number of wires and cells and overall less effective optimization.

\begin{center}
    \begin{tcolorbox}[colback=gray!10,
                      colframe=black,
                      width=\linewidth,
                      arc=1mm, auto outer arc,
                      boxrule=0.5pt,
                     ]
    \textbf{Overall.} These findings highlight the advantage of LLM-based and AST-driven optimizers over traditional EDA tools in data path optimization scenarios. However, their performance on mutant RTL code still lags behind the original, largely due to challenges in fully recognizing and removing functionally redundant or non-executing segments without explicit guidance.
    \end{tcolorbox}
\end{center}


\subsection{Answer to RQ3}

\begin{figure}[!t]
    \centering
     \includegraphics[width=\linewidth]{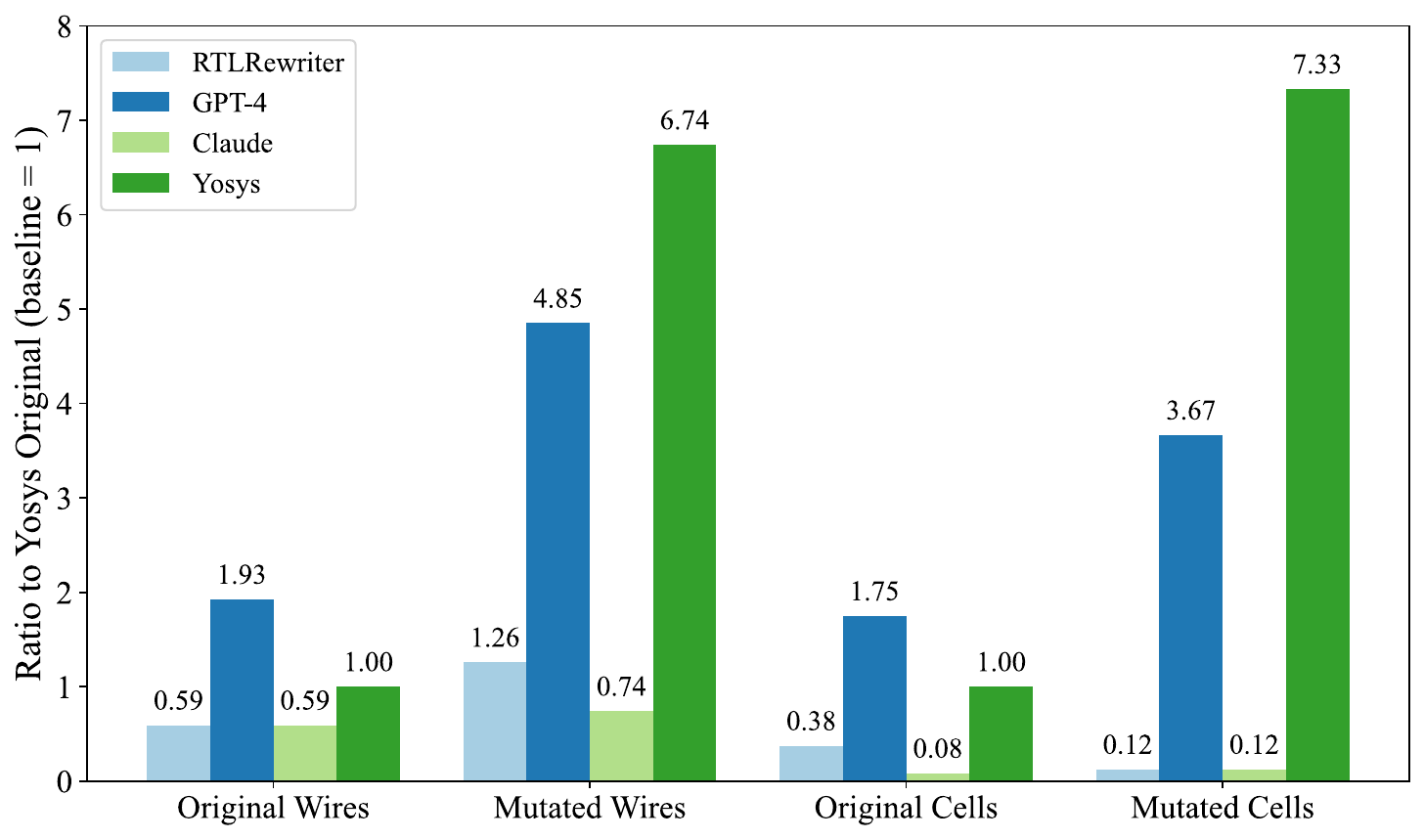}
    \caption{Timing Control Flow Optimization Results with Wires and Cells}
    \label{fig:exp3}
\end{figure}

The results in Table~\ref{tab:timing_control_flow_optimization_results} and \figurename~\ref{fig:exp3} comprehensively illustrate the impact of code mutation on optimization effectiveness. For each method, we observed that the relative ratios for mutant RTL code are consistently worse than those for the original RTL, regardless of the specific metric considered. The phenomenon is likely because code mutation introduces new intermediate states and signal paths that disrupt common optimization heuristics, such as FSM state minimization and redundant logic pruning, which are typically applied in a straightforward control-flow context.

\textbf{Claude-3.7-sonnet} demonstrated the best overall performance by effectively optimizing complex and redundant timing logic. In particular, Claude maintained both area (0.55× baseline) and delay (1.00×) close to those of the original RTL on the mutant RTL code. However, it did not fully eliminate redundant FSM states and only performed partial merging, which limited further improvements in delay.

The performance of \textbf{RTLRewriter} in timing control flow optimization was only better than GPT-4 and Yosys. RTLRewriter retained redundant states in the FSM of the mutant RTL code, resulting in the highest delay among all methods (1.01× baseline) and only modest improvements in area (1.02×). It is primarily due to the partitioning approach adopted by RTLRewriter. Specifically, RTLRewriter would treat the FSM module which contains the segment timing logic as a single submodule for processing. So, it can not explicitly analyze the duration of the redundant state or attempt to shorten state delay chains through techniques like FSM re-encoding or state flattening. These operations require support from the State Transition Graph (STG), which RTLRewriter does not systematically construct.

\begin{table}
    \caption{Timing Control Flow Optimization Results (normalized to Yosys\_org)}
    \centering
    \begin{tabular}{lccc}
        \toprule
        \textbf{Method} & \textbf{Delay Ratio} & \textbf{Area Ratio} & \textbf{Power Ratio} \\
        \midrule
        GPT\_mut          & \cellcolor{mygreen}0.86 $\downarrow$  & \cellcolor{myred}7.93 $\uparrow$ & \cellcolor{myred}3.00 $\uparrow$ \\
        Claude\_mut       & 1.00 --                     & \cellcolor{mygreen}0.55 $\downarrow$  & 1.00 -- \\
        RTLRewriter\_mut  & \cellcolor{myred}1.01 $\uparrow$   & 1.02 $\uparrow$                    & 1.00 -- \\
        Yosys\_mut        & \cellcolor{myred}1.01 $\uparrow$   & 2.93 $\uparrow$                    & 1.00 -- \\
        GPT\_org          & 0.91 $\downarrow$           & 1.78 $\uparrow$                    & 1.00 -- \\
        Claude\_org       & 1.00 --                     & 1.00 --                    & 1.00 -- \\
        RTLRewriter\_org  & 1.00 --                     & 0.61 $\downarrow$                    & 1.00 -- \\
        Yosys\_org        & 1.00 --                     & 1.00 --                    & 1.00 -- \\
        \bottomrule
    \end{tabular}
    \label{tab:timing_control_flow_optimization_results}
\end{table}

Additionally, \textbf{GPT-4} achieved the lowest delay (0.86× baseline) when handling mutant RTL code. This is primarily because, after the timing logic of the RTL code was made more complex, GPT-4 tended to insert registers (a type of cell) into the combinational logic paths. This operation divided the FSM state into multiple shorter segment states rather than combining the redundant states. By doing so, GPT-4 reduced the delay of each segment. But it significantly increased the number of wires (4.85× baseline) and cells (3.67×), as well as power consumption (3.00×). As a result, GPT-4 exhibited low delay but very high area and power overheads when optimizing the mutant RTL code.

\textbf{Yosys} exhibited clear limitations in handling complex timing control flow, with the number of wires and cells increasing by up to 6.74× and 7.33×, respectively, on mutant RTL code.It was less effective at optimizing the redundant state transitions introduced by mutation, resulting in the lowest overall performance among all evaluated methods.

\begin{center}
    \begin{tcolorbox}[colback=gray!10,
                      colframe=black,
                      width=\linewidth,
                      arc=1mm, auto outer arc,
                      boxrule=0.5pt,
                     ]
    \textbf{Overall.} The results indicate that LLM-Based RTL optimization methods are not effective enough in optimizing timing control flow.
    These methods did not eliminate redundant FSM states or thoroughly optimize the complex timing logic.
    \end{tcolorbox}

\end{center}

\subsection{Answer to RQ4}

\begin{figure}[!t]
\centering
 \includegraphics[width=\linewidth]{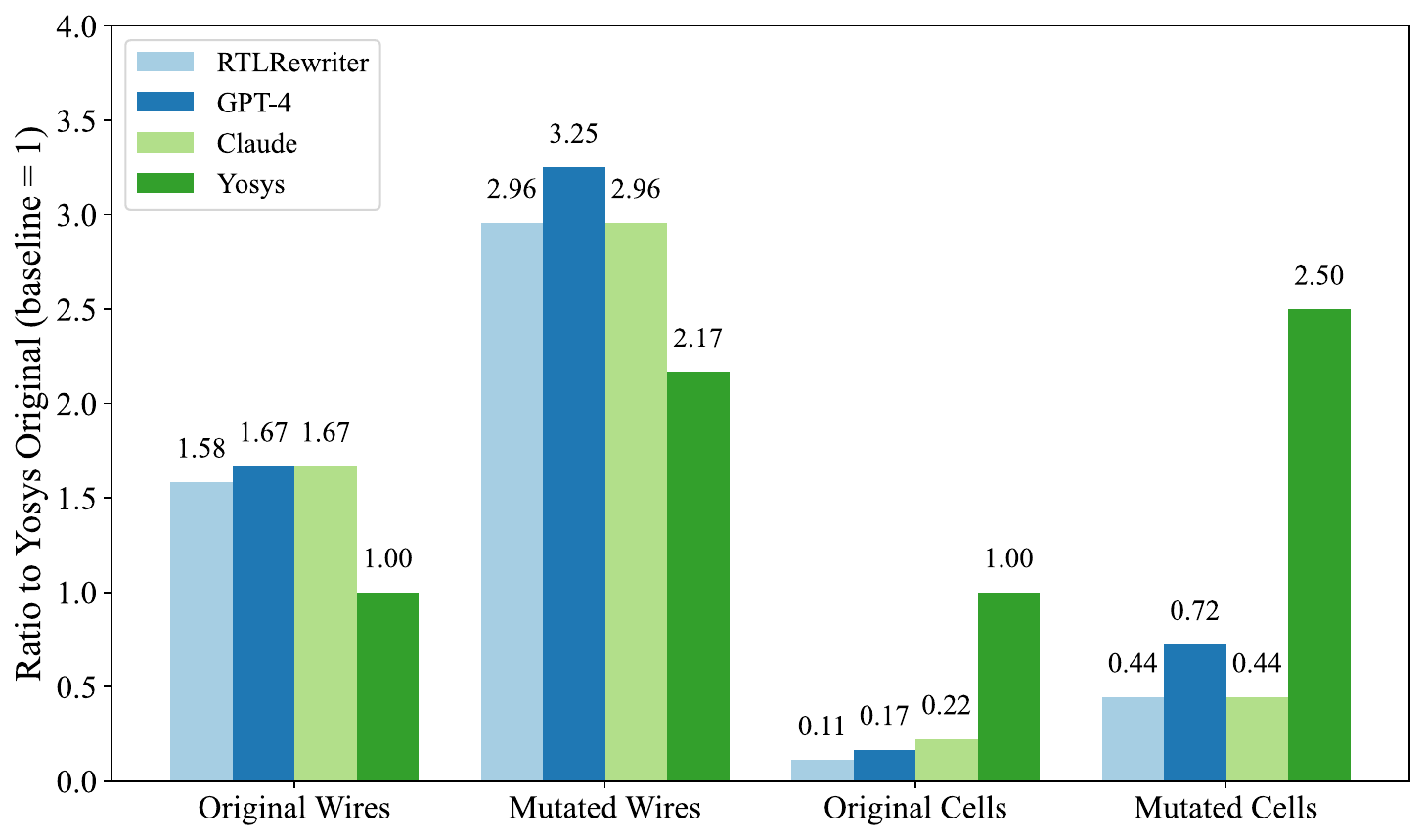}
\caption{Clock Domain Optimization Results with Wires and Cells}
\label{fig:exp4}
\end{figure}

The results in Table~\ref{tab:clock_domain_optimization_results} and \figurename~\ref{fig:exp4} demonstrate that all evaluated methods exhibit diminished optimization effectiveness on mutant RTL with complex clock domains, as compared to their performance on the original RTL code.

\textbf{Claude-3.7-sonnet} again delivered the best overall performance among the tested methods. It achieved delay and area ratios of 1.03 and 1.02 on the mutant RTL code, while maintaining values close to baseline for the original code (0.96 for both metrics). Nevertheless, even Claude is unable to fully optimize the mutant RTL code. For instance, its number of wires and cells remained elevated (1.67× and 0.44× baseline, respectively) on the mutant code, compared to the original. Notably, both \textbf{Claude} and \textbf{GPT-4} tend to emphasize intra-cycle optimizations and do not perform comprehensive inter-cycle or global scheduling adjustments. Therefore, compared to RTLRewriter, both GPT and Claude increased the number of wires and cells on the mutant RTL code(3.25× and 0.72×).

\textbf{RTLRewriter} stood out for reducing the dynamic power consumption on the mutant RTL code (0.77× baseline). It is primarily caused by partitioning the RTL code and lowering the operating clock frequency in each segment, such as through clock gating. However, this approach came at the expense of increased delay (1.09×) and area (1.64×). Because, a lower clock frequency necessitates longer execution times and additional logic resources to manage the segmentation. 
It also highlights one of the disadvantages of the partitioning strategy. The strategy only focuses on optimization within individual segments and cannot effectively perform global scheduling or coordination across the entire RTL code.

\textbf{Yosys}, in contrast, achieved a relatively balanced trade-off between delay (1.07×) and area (1.04×) on the mutant RTL code, without drastic changes to any single metric. However, its overall optimization effectiveness still declined for the mutant code, as indicated by the increased number of wires (2.17×) and cells (2.50×) compared to the original RTL.

\begin{table}
    \caption{Clock Domain Optimization Results (normalized to Yosys\_org)}
    \centering
    \begin{tabular}{lccc}
        \toprule
        \textbf{Method}     & \textbf{Delay Ratio}      & \textbf{Area Ratio}       & \textbf{Power Ratio}   \\
        \midrule
        GPT\_mut            & 1.06 $\uparrow$           & 1.02 $\uparrow$           & \cellcolor{myred}1.00 --   \\
        Claude\_mut         & \cellcolor{mygreen}1.03 $\uparrow$ & \cellcolor{mygreen}1.02 $\uparrow$ & 1.00 --   \\
        RTLRewriter\_mut    & \cellcolor{myred}1.09 $\uparrow$ & \cellcolor{myred}1.64 $\uparrow$ & \cellcolor{mygreen}0.77 $\downarrow$ \\
        Yosys\_mut          & 1.07 $\uparrow$           & 1.04 $\uparrow$           & 1.00 --   \\
        GPT\_org            & 1.00 $\downarrow$         & 0.98 $\downarrow$         & 1.00 --   \\
        Claude\_org         & \cellcolor{mygreen}0.96 $\downarrow$ & \cellcolor{mygreen}0.96 $\downarrow$ & 1.00 --   \\
        RTLRewriter\_org    & \cellcolor{myred}1.01 $\uparrow$ & \cellcolor{myred}1.50 $\uparrow$ & \cellcolor{mygreen}0.62 $\downarrow$ \\
        Yosys\_org          & 1.00 --                  & 1.00 --                  & 1.00 --   \\
        \bottomrule
    \end{tabular}
    \label{tab:clock_domain_optimization_results}
\end{table}

\begin{center}
    \begin{tcolorbox}[colback=gray!10,
                      colframe=black,
                      width=\linewidth,
                      arc=1mm, auto outer arc,
                      boxrule=0.5pt,
                     ]
    \textbf{Overall.}
    These findings indicate that LLM-based RTL optimization methods currently struggle to handle the complexities introduced by structural mutations in clock-domain logic. This limitation is attributable to the inadequate temporal reasoning capabilities of existing LLMs
    \end{tcolorbox}
\end{center}

\section{Discussion}

\subsection{Limitations of LLM-Based RTL Code Optimization Methods}

Although LLM-Based RTL code optimization methods had shown promising results in optimizing RTL code, they still face several limitations. The primary limitation comes from the insufficient understanding of the semantics of the timing logic in RTL code. 
RTL, as a hardware description language, exhibits inherent concurrency, precise timing, and strict dependence on clock cycles, which significantly differ from the sequential text and software code-based training data and processing mechanisms of LLMs~\cite{vericoder}.
Furthermore, existing studies have demonstrated the general limitations of LLMs in broader temporal reasoning capabilities. For example, LLMs struggle with Allen interval relations, causal relationships, and long-term dependencies~\cite{chronosense}. 
The inherent difficulty in interpreting clock-cycle semantics amplifies the typical black-box problem associated with LLMs. Consequently, this can lead to unreliable and non-explainable results during the process of assisting RTL code optimization~\cite{symrtlo}.
But, this is not an unsolvable problem. Helping LLMs better understand the timing information in RTL code could be one of the solutions. For instance, in clock domain optimization, if the prompt includes detailed timing logic of the RTL and explicitly suggests that the LLM merge clock domains, the LLM could achieve better results. Therefore,  
future work could explore the use of temporally-aware neural networks, such as Spiking Neural Network~\cite{SNN,spikingbert}, as a front-end encoder for LLMs to help LLMs better understand the timing logic of RTL code.

\subsection{Threats}

This paper has two main threats to validity. First, the threats in benchmark construction. The quality of the RTL code could affect LLM-Based RTL code optimization. To alleviate this threat, we carefully checked the RTL code in our Benchmark and adjusted the RTL code which is not representative of the real-word RTL code. Second, the generalizability to other LLMs. In this study, we only studied two LLMs and one open-source LLM-Based method due to time and hardware limits. However, there are many other LLMs that could be used for RTL code optimization. To alleviate this threat, we selected the SOTA LLMs in different families as representatives.
Furthermore, there are some LLM-Based methods that can be used for RTL code optimization, such as VeriGen~\cite{vgen} and RTLCoder~\cite{rcoder}. But due to the fact that they mainly focus on RTL code generation rather than optimization, we did not include them in our experiments. 

\subsection{Prompt discussion}
\label{sec:threats}
We acknowledge that the performance of LLM-based RTL optimization is sensitive to prompt design. Different phrasings, examples, or few-shot demonstrations may lead to varied results, potentially affecting both correctness and optimization quality.
To mitigate this, we adopt a fixed zero-shot prompt across all models and mutation scenarios, following the baseline method’s approach and ensuring a controlled evaluation environment. This design choice allows us to isolate the model’s inherent reasoning capability from the confounding influence of prompt engineering.
In future work, we also plan to systematically explore the effect of prompt tuning, including adaptive prompt selection, few-shot exemplars, and reinforcement-guided prompting, to better understand the full potential of LLMs in RTL optimization.

\section{Related Work}

\subsection{Code Optimization}

Code optimization serves as a foundation for various software engineering (SE) tasks~\cite{R1}. Code optimization involves transforming programs at various levels, such as source code~\cite{R2}, compiler intermediate representation~\cite{R3,R12}, or binary code~\cite{R4,R5,R6,R11} to achieve specific performance goals, such as reducing execution time~\cite{R7}, minimizing code size~\cite{R8}, or optimizing memory usage~\cite{R9}. One key challenge in code optimization is the enormous search space, where effective solutions are rare and exhaustive search is computationally impractical~\cite{R41}. Traditionally, code optimization has relied on expert-designed heuristics and rules~\cite{R10,R13}. Over time, a diverse array of optimization techniques has emerged, ranging from low-level strategies, such as instruction scheduling~\cite{R14}, register allocation~\cite{R16}, vectorization~\cite{R42}, and loop transformations~\cite{R43} to high-level strategies that modify algorithms or data structures at the source code level to improve performance~\cite{R15}. 
Over the past decades, a substantial body of work has explored the use of machine learning for code optimization. For example, Liang et al. used deep reinforcement learning to discover pass sequences that reduce both runtime and code size, outperforming standard compiler pipelines for certain benchmarks~\cite{R17}. Cummins et al. proposed ProGraML, it learns graph embeddings of LLVM IR and has been applied to predict optimal optimization passes or to classify programs by optimization difficulty~\cite{R18}.
With the gradual maturity of large language models (LLMs), their application in code optimization has begun to gain attention.
Although LLM-Based code optimization is still at an early stage, some results show that LLMs can occasionally suggest high-level changes, such as adopting alternative algorithms or libraries~\cite{R19}. Nevertheless, LLM-driven code optimization remains immature and faces substantial limitations. Therefore, systematic evaluation methods are needed to assess the effectiveness of LLM-Based code optimization methods.

\subsection{RTL Code Optimization}

Register Transfer Level (RTL) code optimization is crucial for improving the efficiency and performance of digital circuits during the early synthesis stages. In the domain of RTL code optimization, the exploration spans several critical dimensions. For logic operation optimization, researchers focus on subexpression elimination~\cite{R20,R21} , constant folding~\cite{R21}, constant propagation~\cite{R22,R23} and algebraic simplification~\cite{R24,R25}. Concerning data-path optimization, techniques mainly include dead code elimination~\cite{R26,R27}, strength reduction~\cite{R28} Mux reduction~\cite{R29,R30}, Mux tree decomposition~\cite{R31}, and Mux tree restructuring~\cite{R30}. When focusing on memory optimization, the emphasis is  memory sharing~\cite{R32,R33}, memory folding, memory banking~\cite{R34,R35}, and memory pipelining~\cite{R36}. Finally, within finite state machine (FSM) design optimizations, the primary focus is on state minimization~\cite{R37}, state assignment~\cite{R38}, and state decomposition~\cite{R39}. 
Although these techniques are effective, the lack of a centralized repository means that optimization patterns in design manuals and code libraries remain underutilized. This forces engineers to rely more on their expertise rather than on these automated tools.
Recent studies have demonstrated the potential of leveraging the powerful generative capabilities of LLMs to automatically rewrite and optimize RTL code~\cite{RTLRewriter,verigen,rcoder}. For example, Yao et al. proposed RTLRewriter, a method that uses LLMs to rewrite RTL code based on abstract syntax trees (ASTs) and domain-specific knowledge~\cite{RTLRewriter}. It can effectively optimize logic operations, data paths, and timing control flow in RTL code.
However, these methods still face several limitations, such as the inability to align their outputs with specified optimization objectives, inefficiencies in the synthesis process, and lack of timing logic optimization. Therefore, a comprehensive evaluation method is needed to systematically assess the effectiveness of LLM-Based  RTL code rewriting and optimization methods.

\section{Conclusion and Future Work}

In this paper, we propose a new benchmark and a metamorphosis-based evaluation method for evaluating LLM-Based RTL code optimization methods. Intensive experiments are conducted include \textit{Logic operation},\textit{ data path}, \textit{timing control flow} and \textit{timing domain} in RTL code optimization. The results indicate that although LLM-Based RTL code optimization methods can effectively optimize RTL code with complex logic operation and data path. 
It still struggles with RTL code that contains complex timing logic.
In future work, we will focus on improving the LLM-Based RTL code optimization methods to better understand the timing logic of RTL code. By integrating the clock-cycle semantics from static RTL descriptions, LLMs can achieve better optimization results. Additionally, we will explore the integration of domain-specific knowledge into the prompt to enhance the ability of LLMs to optimize RTL code effectively.

\newpage

\bibliographystyle{IEEEtran}
\bibliography{main}

\begin{thebibliography}{10}
\providecommand{\url}[1]{#1}
\csname url@samestyle\endcsname
\providecommand{\newblock}{\relax}
\providecommand{\bibinfo}[2]{#2}
\providecommand{\BIBentrySTDinterwordspacing}{\spaceskip=0pt\relax}
\providecommand{\BIBentryALTinterwordstretchfactor}{4}
\providecommand{\BIBentryALTinterwordspacing}{\spaceskip=\fontdimen2\font plus
\BIBentryALTinterwordstretchfactor\fontdimen3\font minus \fontdimen4\font\relax}
\providecommand{\BIBforeignlanguage}[2]{{%
\expandafter\ifx\csname l@#1\endcsname\relax
\typeout{** WARNING: IEEEtran.bst: No hyphenation pattern has been}%
\typeout{** loaded for the language `#1'. Using the pattern for}%
\typeout{** the default language instead.}%
\else
\language=\csname l@#1\endcsname
\fi
#2}}
\providecommand{\BIBdecl}{\relax}
\BIBdecl

\bibitem{R37}
M.~Harris, ``Synthesis of finite state machines: Functional optimization,'' \emph{Microelectronics Journal}, vol.~29, no.~6, pp. 364--365, 1998.

\bibitem{R38}
T.~Villa, T.~Kam, R.~K. Brayton, and A.~L. Sangiovanni-Vincentelli, \emph{Synthesis of finite state machines: logic optimization}.\hskip 1em plus 0.5em minus 0.4em\relax Springer Science \& Business Media, 2012.

\bibitem{R39}
R.~Shelar, M.~Desai, and H.~Narayanan, ``Decomposition of finite state machines for area, delay minimization,'' in \emph{Proceedings 1999 IEEE International Conference on Computer Design: VLSI in Computers and Processors (Cat. No.99CB37040)}, 1999, pp. 620--625.

\bibitem{R41}
P.~M. Phothilimthana, A.~Thakur, R.~Bodik, and D.~Dhurjati, ``Scaling up superoptimization,'' in \emph{Proceedings of the Twenty-First International Conference on Architectural Support for Programming Languages and Operating Systems}, 2016, pp. 297--310.

\bibitem{RTLRewriter}
X.~Yao, Y.~Wang, X.~Li, Y.~Lian, R.~Chen, L.~Chen, M.~Yuan, H.~Xu, and B.~Yu, ``Rtlrewriter: Methodologies for large models aided rtl code optimization,'' in \emph{Proceedings of the 43rd IEEE/ACM International Conference on Computer-Aided Design}, 2024, pp. 1--7.

\bibitem{symrtlo}
Y.~Wang, W.~Ye, P.~Guo, Y.~He, Z.~Wang, B.~Tian, S.~He, G.~Sun, Z.~Shen, S.~Chen \emph{et~al.}, ``Symrtlo: Enhancing rtl code optimization with llms and neuron-inspired symbolic reasoning,'' \emph{arXiv preprint arXiv:2504.10369}, 2025.

\bibitem{yosys}
C.~Wolf, J.~Glaser, and J.~Kepler, ``Yosys-a free verilog synthesis suite,'' in \emph{Proceedings of the 21st Austrian Workshop on Microelectronics (Austrochip)}, vol.~97, 2013.

\bibitem{R36}
J.~Park and P.~C. Diniz, ``Synthesis of pipelined memory access controllers for streamed data applications on fpga-based computing engines,'' in \emph{Proceedings of the 14th international symposium on Systems synthesis}, 2001, pp. 221--226.

\bibitem{Veval}
M.~Liu, N.~Pinckney, B.~Khailany, and H.~Ren, ``Verilogeval: Evaluating large language models for verilog code generation,'' in \emph{2023 IEEE/ACM International Conference on Computer Aided Design (ICCAD)}.\hskip 1em plus 0.5em minus 0.4em\relax IEEE, 2023, pp. 1--8.

\bibitem{rtllm}
Y.~Lu, S.~Liu, Q.~Zhang, and Z.~Xie, ``Rtllm: An open-source benchmark for design rtl generation with large language model,'' in \emph{2024 29th Asia and South Pacific Design Automation Conference (ASP-DAC)}.\hskip 1em plus 0.5em minus 0.4em\relax IEEE, 2024, pp. 722--727.

\bibitem{chen2009design}
D.~Chen, ``, design automation for microelectronics, springer handbook of automation,'' \emph{icims. csl. uiuc. edu}, 2009.

\bibitem{sasao1993logic}
T.~Sasao, \emph{Logic synthesis and optimization}.\hskip 1em plus 0.5em minus 0.4em\relax Springer, 1993, vol.~2.

\bibitem{zimmermann2005rtl}
R.~Zimmermann and A.~Syed, ``Rtl coding guidelines for datapath synthesis,'' Presentation, Synopsys Users Group (SNUG) Boston, 2005, \url{https://picture.iczhiku.com/resource/eetop/wyiEELySpYPudnxV.pdf}.

\bibitem{synopsys2011dc}
{Synopsys Inc.}, \emph{Design Compiler Optimization Reference Manual}, vf-2011.09~ed., Synopsys, 2011, \url{https://picture.iczhiku.com/resource/eetop/SHidRGQWtQruovNN.pdf}.

\bibitem{doulos2020fsm}
{Doulos Ltd.}, ``{FSM Optimization: KnowHow Tutorial},'' \url{https://www.doulos.com/knowhow/fpga/fsm-optimization/}, 2020, accessed: 2025-07-22.

\bibitem{amd2023ug949}
{AMD (Xilinx)}, \emph{{Vivado Design Suite User Guide: Synthesis and Timing Closure (UG949)}}, \url{https://docs.amd.com/r/2023.2-English/ug949-vivado-design-methodology}, 2023, accessed: 2025-07-22.

\bibitem{cadence2014cdc}
{Cadence Design Systems}, ``Clock-domain and reset-domain crossing in low-power design,'' \emph{Tech Design Forum}, 2014, accessed: 2025-07-22.

\bibitem{vemeko2025fpga}
{Vemeko Blog}, ``What techniques can be used to reduce power consumption in fpgas?'' \url{https://www.vemeko.com/blog/67173.html}, Apr. 2025, accessed: 2025-07-22.

\bibitem{unnikrishnan2024rtl}
S.~Unnikrishnan and S.~M. Iype. (2024, Feb.) Reducing power hot spots through rtl optimization techniques. \url{https://www.design-reuse.com/article/61502-reducing-power-hot-spots-through-rtl-optimization-techniques/}. Accessed: 2025-07-22.

\bibitem{SIMTAM}
Z.~Xu, S.~Guo, X.~Li, Z.~Wang, and H.~Jiang, ``Simtam: Generation diversity test programs for fpga simulation tools testing via timing area mutation,'' \emph{ACM Transactions on Design Automation of Electronic Systems}, vol.~30, no.~2, pp. 1--25, 2025.

\bibitem{LegoHDL}
Z.~Xu, S.~Guo, G.~Zhao, P.~Zou, X.~Li, and H.~Jiang, ``A novel hdl code generator for effectively testing fpga logic synthesis compilers,'' \emph{IEEE Transactions on Computer-Aided Design of Integrated Circuits and Systems}, 2025.

\bibitem{verilogeval}
M.~Liu, N.~Pinckney, B.~Khailany, and H.~Ren, ``Verilogeval: Evaluating large language models for verilog code generation,'' in \emph{2023 IEEE/ACM International Conference on Computer Aided Design (ICCAD)}.\hskip 1em plus 0.5em minus 0.4em\relax IEEE, 2023, pp. 1--8.

\bibitem{verigen}
S.~Thakur, B.~Ahmad, H.~Pearce, B.~Tan, B.~Dolan-Gavitt, R.~Karri, and S.~Garg, ``Verigen: A large language model for verilog code generation,'' \emph{ACM Transactions on Design Automation of Electronic Systems}, vol.~29, no.~3, pp. 1--31, 2024.

\bibitem{rtlfixer}
Y.~Tsai, M.~Liu, and H.~Ren, ``Rtlfixer: Automatically fixing rtl syntax errors with large language model,'' in \emph{Proceedings of the 61st ACM/IEEE Design Automation Conference}, 2024, pp. 1--6.

\bibitem{Pei}
K.~Pei, W.~Li, Q.~Jin, S.~Liu, S.~Geng, L.~Cavallaro, J.~Yang, and S.~Jana, ``Exploiting code symmetries for learning program semantics,'' in \emph{Forty-first International Conference on Machine Learning, {ICML} 2024, Vienna, Austria, July 21-27, 2024}, 2024.

\bibitem{R46}
X.~Jin, K.~Pei, J.~Y. Won, and Z.~Lin, ``Symlm: Predicting function names in stripped binaries via context-sensitive execution-aware code embeddings,'' in \emph{Proceedings of the 2022 ACM SIGSAC Conference on Computer and Communications Security}, 2022, pp. 1631--1645.

\bibitem{codellama}
B.~Roziere, J.~Gehring, F.~Gloeckle, S.~Sootla, I.~Gat, X.~E. Tan, Y.~Adi, J.~Liu, R.~Sauvestre, T.~Remez \emph{et~al.}, ``Code llama: Open foundation models for code,'' \emph{arXiv preprint arXiv:2308.12950}, 2023.

\bibitem{R44}
J.~Henkel, G.~Ramakrishnan, Z.~Wang, A.~Albarghouthi, S.~Jha, and T.~Reps, ``Semantic robustness of models of source code,'' in \emph{2022 IEEE International Conference on Software Analysis, Evolution and Reengineering (SANER)}.\hskip 1em plus 0.5em minus 0.4em\relax IEEE, 2022, pp. 526--537.

\bibitem{R45}
S.~Ullah, M.~Han, S.~Pujar, H.~Pearce, A.~Coskun, and G.~Stringhini, ``Can large language models identify and reason about security vulnerabilities? not yet,'' \emph{arXiv preprint arXiv:2312.12575}, 2023.

\bibitem{Iverilog}
``Icarus verilog,'' \url{https://github.com/steveicarus/iverilog}, 2023.

\bibitem{vericoder}
A.~Wei, H.~Tan, T.~Suresh, D.~Mendoza, T.~S. Teixeira, K.~Wang, C.~Trippel, and A.~Aiken, ``Vericoder: Enhancing llm-based rtl code generation through functional correctness validation,'' \emph{arXiv preprint arXiv:2504.15659}, 2025.

\bibitem{chronosense}
D.~S. Islakoglu and J.-C. Kalo, ``Chronosense: Exploring temporal understanding in large language models with time intervals of events,'' \emph{arXiv preprint arXiv:2501.03040}, 2025.

\bibitem{SNN}
S.~Ghosh-Dastidar and H.~Adeli, ``Spiking neural networks,'' \emph{International journal of neural systems}, vol.~19, no.~04, pp. 295--308, 2009.

\bibitem{spikingbert}
M.~Bal and A.~Sengupta, ``Spikingbert: Distilling bert to train spiking language models using implicit differentiation,'' in \emph{Proceedings of the AAAI conference on artificial intelligence}, vol.~38, no.~10, 2024, pp. 10\,998--11\,006.

\bibitem{vgen}
S.~Thakur, B.~Ahmad, H.~Pearce, B.~Tan, B.~Dolan-Gavitt, R.~Karri, and S.~Garg, ``Verigen: A large language model for verilog code generation,'' \emph{ACM Transactions on Design Automation of Electronic Systems}, vol.~29, no.~3, pp. 1--31, 2024.

\bibitem{rcoder}
S.~Liu, W.~Fang, Y.~Lu, J.~Wang, Q.~Zhang, H.~Zhang, and Z.~Xie, ``Rtlcoder: Fully open-source and efficient llm-assisted rtl code generation technique,'' \emph{IEEE Transactions on Computer-Aided Design of Integrated Circuits and Systems}, 2024.

\bibitem{R1}
E.~S. Lowry and C.~W. Medlock, ``Object code optimization,'' \emph{Communications of the ACM}, vol.~12, no.~1, pp. 13--22, 1969.

\bibitem{R2}
A.~Shypula, A.~Madaan, Y.~Zeng, U.~Alon, J.~Gardner, M.~Hashemi, G.~Neubig, P.~Ranganathan, O.~Bastani, and A.~Yazdanbakhsh, ``Learning performance-improving code edits,'' \emph{arXiv preprint arXiv:2302.07867}, 2023.

\bibitem{R3}
C.~Cummins, V.~Seeker, D.~Grubisic, B.~Roziere, J.~Gehring, G.~Synnaeve, and H.~Leather, ``Meta large language model compiler: Foundation models of compiler optimization,'' \emph{arXiv preprint arXiv:2407.02524}, 2024.

\bibitem{R12}
M.~Willsey, C.~Nandi, Y.~R. Wang, O.~Flatt, Z.~Tatlock, and P.~Panchekha, ``Egg: Fast and extensible equality saturation,'' \emph{Proceedings of the ACM on Programming Languages}, vol.~5, no. POPL, pp. 1--29, 2021.

\bibitem{R4}
M.~A. Ben~Khadra, D.~Stoffel, and W.~Kunz, ``Efficient binary-level coverage analysis,'' in \emph{Proceedings of the 28th ACM Joint Meeting on European Software Engineering Conference and Symposium on the Foundations of Software Engineering}, 2020, pp. 1153--1164.

\bibitem{R5}
M.~F. Fernandez, ``Simple and effective link-time optimization of modula-3 programs,'' in \emph{Proceedings of the ACM SIGPLAN 1995 conference on Programming language design and implementation}, 1995, pp. 103--115.

\bibitem{R6}
N.~Licker and T.~M. Jones, ``Duplo: a framework for ocaml post-link optimisation,'' \emph{Proceedings of the ACM on Programming Languages}, vol.~4, no. ICFP, pp. 1--29, 2020.

\bibitem{R11}
A.~A. Moreira, G.~Ottoni, and F.~M. Quint{\~a}o~Pereira, ``Vespa: static profiling for binary optimization,'' \emph{Proceedings of the ACM on Programming Languages}, vol.~5, no. OOPSLA, pp. 1--28, 2021.

\bibitem{R7}
A.~Madaan, N.~Tandon, P.~Gupta, S.~Hallinan, L.~Gao, S.~Wiegreffe, U.~Alon, N.~Dziri, S.~Prabhumoye, Y.~Yang \emph{et~al.}, ``Self-refine: Iterative refinement with self-feedback,'' \emph{Advances in Neural Information Processing Systems}, vol.~36, pp. 46\,534--46\,594, 2023.

\bibitem{R8}
D.~Grubisic, V.~Seeker, G.~Synnaeve, H.~Leather, J.~Mellor-Crummey, and C.~Cummins, ``Priority sampling of large language models for compilers,'' in \emph{Proceedings of the 4th Workshop on Machine Learning and Systems}, 2024, pp. 91--97.

\bibitem{R9}
S.~Garg, R.~Z. Moghaddam, C.~B. Clement, N.~Sundaresan, and C.~Wu, ``Deepdev-perf: a deep learning-based approach for improving software performance,'' in \emph{Proceedings of the 30th ACM Joint European Software Engineering Conference and Symposium on the Foundations of Software Engineering}, 2022, pp. 948--958.

\bibitem{R10}
A.~K. Sarma, ``New trends and challenges in source code optimization,'' \emph{International Journal of Computer Applications}, vol. 131, no.~16, pp. 27--32, 2015.

\bibitem{R13}
A.~Lepori, A.~Calotoiu, and T.~Hoefler, ``Iterating pointers: Enabling static analysis for loop-based pointers,'' \emph{ACM Transactions on Architecture and Code Optimization}, vol.~22, no.~1, pp. 1--25, 2025.

\bibitem{R14}
O.~Fl{\"u}ckiger, G.~Chari, M.-H. Yee, J.~Je{\v{c}}men, J.~Hain, and J.~Vitek, ``Contextual dispatch for function specialization,'' \emph{Proceedings of the ACM on Programming Languages}, vol.~4, no. OOPSLA, pp. 1--24, 2020.

\bibitem{R16}
G.~J. Chaitin, ``Register allocation \& spilling via graph coloring,'' \emph{ACM Sigplan Notices}, vol.~17, no.~6, pp. 98--101, 1982.

\bibitem{R42}
Y.~Chen, C.~Mendis, and S.~Amarasinghe, ``All you need is superword-level parallelism: systematic control-flow vectorization with slp,'' in \emph{Proceedings of the 43rd ACM SIGPLAN International Conference on Programming Language Design and Implementation}, 2022, pp. 301--315.

\bibitem{R43}
A.~Venkat, M.~Hall, and M.~Strout, ``Loop and data transformations for sparse matrix code,'' \emph{ACM SIGPLAN Notices}, vol.~50, no.~6, pp. 521--532, 2015.

\bibitem{R15}
M.~M. Vitousek, J.~G. Siek, and A.~Chaudhuri, ``Optimizing and evaluating transient gradual typing,'' in \emph{Proceedings of the 15th ACM SIGPLAN international symposium on dynamic languages}, 2019, pp. 28--41.

\bibitem{R17}
Y.~Liang, K.~Stone, A.~Shameli, C.~Cummins, M.~Elhoushi, J.~Guo, B.~Steiner, X.~Yang, P.~Xie, H.~J. Leather \emph{et~al.}, ``Learning compiler pass orders using coreset and normalized value prediction,'' in \emph{International Conference on Machine Learning}.\hskip 1em plus 0.5em minus 0.4em\relax PMLR, 2023, pp. 20\,746--20\,762.

\bibitem{R18}
C.~Cummins, Z.~V. Fisches, T.~Ben-Nun, T.~Hoefler, M.~F. O’Boyle, and H.~Leather, ``Programl: A graph-based program representation for data flow analysis and compiler optimizations,'' in \emph{International Conference on Machine Learning}.\hskip 1em plus 0.5em minus 0.4em\relax PMLR, 2021, pp. 2244--2253.

\bibitem{R19}
J.~Gong, V.~Voskanyan, P.~Brookes, F.~Wu, W.~Jie, J.~Xu, R.~Giavrimis, M.~Basios, L.~Kanthan, and Z.~Wang, ``Language models for code optimization: Survey, challenges and future directions,'' \emph{arXiv preprint arXiv:2501.01277}, 2025.

\bibitem{R20}
R.~Pasko, P.~Schaumont, V.~Derudder, S.~Vernalde, and D.~Durackova, ``A new algorithm for elimination of common subexpressions,'' \emph{IEEE Transactions on Computer-Aided Design of Integrated Circuits and Systems}, vol.~18, no.~1, pp. 58--68, 1999.

\bibitem{R21}
J.~Cocke, ``Global common subexpression elimination,'' in \emph{Proceedings of a symposium on Compiler optimization}, 1970, pp. 20--24.

\bibitem{R22}
M.~N. Wegman and F.~K. Zadeck, ``Constant propagation with conditional branches,'' \emph{ACM Transactions on Programming Languages and Systems (TOPLAS)}, vol.~13, no.~2, pp. 181--210, 1991.

\bibitem{R23}
R.~Metzger and S.~Stroud, ``Interprocedural constant propagation: An empirical study,'' \emph{ACM Letters on Programming Languages and Systems (LOPLAS)}, vol.~2, no. 1-4, pp. 213--232, 1993.

\bibitem{R24}
B.~Buchberger and R.~Loos, ``Algebraic simplification,'' in \emph{Computer algebra: symbolic and algebraic computation}.\hskip 1em plus 0.5em minus 0.4em\relax Springer, 1982, pp. 11--43.

\bibitem{R25}
J.~Carette, ``Understanding expression simplification,'' in \emph{Proceedings of the 2004 international symposium on Symbolic and algebraic computation}, 2004, pp. 72--79.

\bibitem{R26}
J.~Knoop, O.~R{\"u}thing, and B.~Steffen, ``Partial dead code elimination,'' \emph{ACM Sigplan Notices}, vol.~29, no.~6, pp. 147--158, 1994.

\bibitem{R27}
R.~Gupta, D.~Benson, and J.~Z. Fang, ``Path profile guided partial dead code elimination using predication,'' in \emph{Proceedings 1997 International Conference on Parallel Architectures and Compilation Techniques}.\hskip 1em plus 0.5em minus 0.4em\relax IEEE, 1997, pp. 102--113.

\bibitem{R28}
K.~D. Cooper, L.~T. Simpson, and C.~A. Vick, ``Operator strength reduction,'' \emph{ACM Transactions on Programming Languages and Systems (TOPLAS)}, vol.~23, no.~5, pp. 603--625, 2001.

\bibitem{R29}
D.~Chen and J.~Cong, ``Register binding and port assignment for multiplexer optimization,'' in \emph{ASP-DAC 2004: Asia and South Pacific Design Automation Conference 2004 (IEEE Cat. No. 04EX753)}.\hskip 1em plus 0.5em minus 0.4em\relax IEEE, 2004, pp. 68--73.

\bibitem{R30}
Z.~Wang, H.~You, J.~Wang, M.~Liu, Y.~Su, and Y.~Zhang, ``Optimization of multiplexer combination in rtl logic synthesis,'' in \emph{2023 International Symposium of Electronics Design Automation (ISEDA)}.\hskip 1em plus 0.5em minus 0.4em\relax IEEE, 2023, pp. 121--125.

\bibitem{R31}
P.~Pi{\v{s}}teka, K.~Jelemensk{\'a}, and M.~Koles{\'a}r, ``Reduction of multiplexer trees using modified lookup table.''

\bibitem{R32}
C.~E. LaForest and J.~G. Steffan, ``Efficient multi-ported memories for fpgas,'' in \emph{Proceedings of the 18th annual ACM/SIGDA international symposium on Field programmable gate arrays}, 2010, pp. 41--50.

\bibitem{R33}
J.~Ma, G.~Zuo, K.~Loughlin, X.~Cheng, Y.~Liu, A.~M. Eneyew, Z.~Qi, and B.~Kasikci, ``A hypervisor for shared-memory fpga platforms,'' in \emph{Proceedings of the Twenty-Fifth International Conference on Architectural Support for Programming Languages and Operating Systems}, 2020, pp. 827--844.

\bibitem{R34}
Y.~Zhou, K.~M. Al-Hawaj, and Z.~Zhang, ``A new approach to automatic memory banking using trace-based address mining,'' in \emph{Proceedings of the 2017 ACM/SIGDA International Symposium on Field-Programmable Gate Arrays}, 2017, pp. 179--188.

\bibitem{R35}
B.-C. Lai, B.-Y. Chen, B.-E. Chen, and Y.-D. Hsin, ``Remap+: An efficient banking architecture for multiple writes of algorithmic memory,'' \emph{IEEE Transactions on Very Large Scale Integration (VLSI) Systems}, vol.~28, no.~3, pp. 660--671, 2019.

\end{thebibliography}

\end{document}